**Title.** Biodiversity data standards for the organization and dissemination of complex research projects and digital twins: a guide

**Authors.** Carrie Andrew[1], Sharif Islam[2], Claus Weiland[3], Dag Endresen[1]

**Institutions.** [1]Natural History Museum, University of Oslo, Sars' gate 1, 0562 Oslo, Norway; [2] Naturalis Biodiversity Center, Darwinweg 2, 2333 CR Leiden, Netherlands; [3] Senckenberg – Leibniz Institution for Biodiversity and Earth System Research, Senckenberganlage 25, 60325 Frankfurt, Germany.

**ORCID's.** CA: https://orcid.org/0000-0002-0524-8334; SI: https://orcid.org/0000-0001-8050-0299; CW: https://orcid.org/0000-0003-0351-6523; DE: https://orcid.org/0000-0002-2352-5497.

**Emails.** CA: c.j.andrew@nhm.uio.no; SI: sharif.islam@naturalis.nl; CW: claus.weiland@senckenberg.de; DE dag.endresen@nhm.uio.no.

**Abstract.** Biodiversity data are substantially increasing, spurred by technological advances and community (citizen) science initiatives. To integrate data is, likewise, becoming more commonplace. Open science promotes open sharing and data usage. Data standardization is an instrument for the organization and integration of biodiversity data, which is required for complex research projects and digital twins. However, just like with an actual instrument, there is a learning curve to understanding the data standards field. Here we provide a guide, for data providers and data users, on the logistics of compiling and utilizing biodiversity data. We emphasize data standards, because they are integral to data integration. Three primary avenues for compiling biodiversity data are compared, explaining the importance of research infrastructures for coordinated long-term data aggregation. We exemplify the Biodiversity Digital Twin (BioDT) as a case study. Four approaches to data standardization are presented in terms of the balance between practical constraints and the advancement of the data standards field. We aim for this paper to guide and raise awareness of the existing issues related to data standardization, and especially how data standards are key to data interoperability, i.e., machine accessibility. The future is promising for computational biodiversity advancements, such as with the BioDT project, but it rests upon the shoulders of machine actionability and readability, and that requires data standards for computational communication.

## 1. Introduction. The need to organize, integrate disseminate and utilize biodiversity data.

Biodiversity data are being generated at an exceedingly rapid rate, propelled by novel techniques (e.g., eDNA, drones, satellites, camera traps, acoustic monitors, text extraction, AI) and community (citizen) science initiatives to gather information at levels never before achieved. It is an exciting time for biodiversity research (Heberling et al. 2021), as the greater extent of data makes unanswered questions in ecology and evolution ever more obtainable. While they can help address them, rarely is it instantaneous. Instead, biodiversity data are a massively increasing resource. They must be organized, and then disseminated, in order to be effectively utilized. There are challenges to integrating biodiversity data, as we describe here.

We explain, step-by-step, the logistics of compiling and utilizing biodiversity data, when considered as a whole across independent data collection events. Our target audiences are biodiversity data providers and biodiversity data users, for whom we find that such a guideline resource is currently lacking, yet would prove helpful to be provided. We aim for readers to find this guide useful as a general reference for understanding integrated biodiversity data, with our final recommendations aimed towards research infrastructures. We begin with the organization of biodiversity data for dissemination, and emphasize data standards, because they are how data become integrated. As will be exemplified, no matter the solution, new challenges will - and do - arise with compiling and integrating "big data". We then utilize the Biodiversity Digital Twin (de Koning et al. 2023, Trantas et al. 2023, Golivets et al. 2024) as a data standards case study, because it exemplifies the varied states of data standards that most biodiversity research projects are involved with, and the needs to support technological advances alongside data mobilization within data infrastructures. We illustrate the processes that are involved to simultaneously advance multiple research projects, each with different data sources and modeling outcomes, to function as one singular biodiversity resource, i.e., a digital twin.

There are two assumptions that we begin with. First, that biodiversity data, across independent research projects, should be made Findable, Accessible, Interoperable and Reusable, i.e., FAIR (Wilkinson et al. 2016) - and not relegated to a trash bin, whether real or digital. Second, that data should be deposited in integrated databases and data infrastructures, instead of private initiatives, to help establish their reusability and preserve their longevity.

**2. From fractionated datasets to unified resources.**

Whether by researchers or community scientists, and whether from a single project or a global initiative, people and machines are creating vast amounts of biodiversity data - in parallel independent projects. This is a challenge, as biodiversity data typically originate from multiple, fractionated sources, and often from dissimilar methodologies and measures (Gadelha et al. 2020). Data also vary in their degrees of FAIRness, which impacts their accessibility for people, as well as their machine interoperability and readability (Wilkinson et al. 2016). Computationally, software agents are able to autonomously interpret and process FAIR data, meaning that they are machine-actionable (Jacobsen et al. 2019; Section 3).

There are three common ways to find data resources and to deal with data integration: Either a) a list of available data must be created, which the community can access for independent integration, or data must be deposited within established b) databases or c) data infrastructures for community access of integrated data (Figure 1). No solution is ideal, but, as we will explain, shared or interoperable data standards makes data fusion significantly easier.

Creating a list of available data sources is the only option that does not directly work with the data, but which does inventory the data which are available. It is, therefore, most useful when data sources still need to be compiled. Take, for example, global conservation initiatives. Inadequate data are a source of frustration that impedes conservation policy and management (Stephenson et al. 2022). One solution has been to compile a list of existing conservation data sources (e.g., Stephenson & Stengel 2020). People can independently access and integrate the data from the list, depending on their own objectives. A list option is impractical across all biodiversity topics, having previously been attempted (Blair et al. 2020), but it can function successfully for subtopics, e.g., conservation (Stephenson & Stengel 2020).

However, lists that have been compiled require consistent updating to remain actively usable, as new datasets may become available and/or existing datasets may become degenerated, especially if not made FAIR (e.g., due to nonfunctional weblinks). Lists may also result in a lot of duplicated effort, since data integration will occur independently between users, and without any of them necessarily openly sharing their harmonized data. It is important to note that the lack of sharing may be due to data protection rights of at least one of the original data sources. In those cases, the data will always need to be individually accessed, irrespective of any data integration and sharing initiatives. It is also not always possible to integrate all of the data listed if the contents do not match. There are reasons, thus, for database lists in certain situations. Applying a community-driven approach to curating and maintaining any list could extend its effective lifespan, as many people would be able to help contribute to and update it (Blair et al. 2020).

It may be more beneficial to, instead, create a database that researchers can submit their data to (Figure 1). This may occur independently through the creation of a single, accessible database. For example, often such a database, begun by a select group of researchers, is first published in a peer-reviewed journal and then opened up for further data accessions to it, as well as use of the data. When accepted by the general scientific community, this is an effective bottom-up approach to unify biodiversity data. For example, this approach has proven popular for traits-based data; both the TRY (Plant trait database; Kattge et al. 2011) and the FRED (Fine-Root Ecology Database, Iversen et al. 2017) databases are examples of such initiatives. However, by being kickstarted via independent researchers &/or within specific institutions, the continual requirements for maintenance of any database may, as with lists, be difficult to perpetuate across longer time scales. Project funded dependency limits the duration as well as the benefits of the aggregated data to the initiative's lifetime. Community-driven approaches again have the potential to maintain databases longer than individual people can (Blair et al. 2020). For example, both the TRY and FRED databases can now be found as part of the Open Traits Network (Gallagher et al. 2020), a community-driven organismal traits database initiative that has unified earlier initiatives into one.

Currently there is hope by some that the management of both lists and databases can be sustained by willing community members (e.g., Blair et al. 2020, Gallagher et al. 2020). However, how durable such distributed collaboration initiatives can be depends on a multitude of factors related to user demographics, degree of support by the hosting institution, and, ultimately, still the need for database management by a select group of people. Lists and databases require maintenance by groups of dedicated individuals, and, in addition, they can be prone to errors from public contributors, if not checked or limited. Community-driven approaches to lists and database are, thus, not automatic promises of success (Shaw & Hargittai 2018, O'Leary et al. 2020).

It may be more sustainable in the long-term to integrate data into a *data infrastructure*, which is the third option (Guralnick et al. 2007). A larger network of support, and longer funding duration of institutionalized, non-project-based data infrastructures, which essentially are libraries for digitized data, should provide protection and access to the data across many generations of scientists. One example is the Global Biodiversity Information Facility, GBIF (Guralnick et al. 2007, Robertson et al. 2014, de Poorter et al. 2017). GBIF has functioned by building upon local data repositories to aggregate data into their infrastructure. Dedicated individuals (nationally funded nodes staff and globally funded secretariat staff) work to maintain and integrate the existing and newly deposited data, while community-driven approaches are also applied, but as subcomponents for data deposition ultimately managed by regulations implemented by GBIF employees (Robertson et al. 2014), and set by the wider governing community TDWG (Section 3).

Data in a data infrastructure must fit into the protocols that have been established. It is a result of what is referred to as a *centralized and connected network architecture design* for a data infrastructure (Gallagher et al. 2020); the biodiversity data are *connected* into a data infrastructure from multiple sources, but are limited to the formatting that a *central* hub has designated, as the manager of the data. The connection to many sources is not the issue with the network design, but the centralization aspect can be seen as a governance issue, despite that such requirements are often imposed due to standardization requirements (discussed further in Sections 3 - 5). Another caveat of data infrastructures is that no single infrastructure can capture all of biodiversity data (De Pooter et al. 2017). And, finally, data infrastructures help streamline data deposition and management, but they, ultimately, function the same as a database to many data depositors and users. They may, thus, end up on a list alongside other individual databases, for any given biodiversity data topic.

Given the need for multiple lists, databases &/or data infrastructures, and given the complexity of biodiversity data, there is the potential for overlap. Databases originally created independently may, with time, be deposited within a data infrastructure, perhaps after it was developed or gained popularity. Or databases may be combined into a larger one (Gallagher et al. 2020). Data infrastructures pull in data from multiple sources, despite repetition, although this is, to a degree, contingent on the taxonomic systems utilized (Feng et al. 2022). Data can, thus, fit within multiple lists, databases and/or data infrastructure categories, which does create data duplication. So long as key information (e.g., a persistent identifier or combination of original source, date, location, taxon, etc.) is passed along to the different resources, duplication is not likely to be a significant issue to data users. Importantly, the key information must be standardized in order to most easily discern any data repetitions between sources.

## 3. There are (many) standards to follow.

As long as the aim is to save and share biodiversity data, we advocate that the most optimal solutions are to deposit data within databases and data infrastructures. However, it requires harmonization and integration of the data, i.e., they need to be standardized (Box 1). Broadly considered, there are multiple types of, and reasons for using, data standards. For example, contents of variables have standards (i.e., units or forms of measurement, such as the SI units), variables need to be standardized by name and meaning across datasets, and even metadata require standards for machine interoperability and readability. For the purposes here, we focus on the standardization of biodiversity datasets, including the relationships between them and the data they contain (e.g., taxonomic hierarchy or interspecies interactions). We utilize the term "dataset record-level data standards" to differentiate them from the other forms.

Standardization can occur very simply and independently, for example by a single researcher integrating datasets, they do employ a form of standardization. Collectively and more formally, standards are ratified at an international level for widespread adoption and utilization. The Biodiversity Information Standards working group (TDWG, the acronym is based on a former name of the group; https://www.tdwg.org) is the international resource for biodiversity data standards. It is a collaborative international network of scientists and computational experts from a variety of fields, who together establish standards for dataset integration. Data standards have developed primarily based on the interest and computational needs of those depositing and integrating data; thus, they have been especially influenced by museum data digitization (e.g., Graham et al. 2004), and the rise of data infrastructures such as GBIF (Wieczorek et al. 2012, Robertson et al. 2014, De Pooter et al. 2017). However, as the field of biodiversity data changes, and the data sources diversify, so too do the practicalities of data standardization (e.g., Schneider et al. 2019; Section 4).

Dataset record-level data standards help integrate and harmonize data into databases, whether private or shared, and, ultimately, into data infrastructures. They make different datasets uniform during integration, by requiring formatting of data to established identifiers (e.g., persistent identifiers such as DOIs), labels (e.g., names of variables), definitions (e.g., how variables are defined) and interrelationships (e.g., how variables interrelate). The focus is, from the data providers' and users' points of view, primarily on the variables: how they are named and how they interrelate (Figure 1). From the data scientists' point of view, it is more complicated, because very large databases do not usually store data as a single table that retains variables as, for example, the columns. They instead parse data to save storage space, often by reducing redundancies and blank entries. For example, they work around *data cores*, which are subsets of the original datasets that connect to all of the rest of the dataset material (and, thus, need to be standardized).

*Data cores* are an example of required (i.e., core) variables for data deposition into a database or data infrastructure. They have become somewhat synonymous with dataset record-level standards because the terms (i.e., names of variable) have established vocabularies, thesauri and/or ontologies associated with them (Box 1). For example, GBIF's star schema is built around the Darwin Core data standard, a set of terms (variable names) describing the taxon, location, date and related basic information, and which have been ratified in TDWG (Wieczorek et al. 2012, Baskauf et al. 2016, Baskauf & Webb 2016). Cores function also to limit the amount of storage space that data take up by revolving around a core standard of terms, compared to if data were stored in a more basic table format - although data cores have been criticized for constraining how easily new variables can be added, thus they may become obsolete with time (Robertson et al. 2014, De Pooter et al. 2017, Gallagher et al. 2020; Section 4 below). There is rarely a perfect match to how a database structure is set up;

tradeoffs between types of data, storage space, computational time to access data, and the incorporation of new forms of data all weigh into any option.

Dataset record-level data standards, more concretely, include collections of terms (e.g., names of variable), associated concepts of the terms (e.g., definitions), and described relationships between the terms (e.g., taxonomic hierarchies or species interactions between variables). They are assembled into *semantic artifacts* such as *vocabularies*, *thesauri* and/or *ontologies*. They often follow semantic web standards via the application of the *resource description framework* (*RDF*), as expressed with the *Web Ontology Language* (*OWL*) for computational relevancy (Box 1). It can be a major semantic endeavor to create any of the types of standards, requiring many participants for discussions, and the general scientific community for implementation. When employed, they define and standardize variables, e.g., the data fields across different data sets, make data integration possible, and can pave the way for *machine actionability* (with a consistent structure for computation) and *machine readability* (with instructions for computational interpretation and manipulation).

Despite their relatedness, there are critical differences between the forms of dataset record-level data standards. *Vocabularies* are lists of words (terms, such as variable names) with definitions, like a dictionary. They organize specific terms (names of variables) into an inventory. For example, the Darwin Core standards list terms and their definitions in a controlled vocabulary. Many times, however, relationships between terms are needed to better express them. A classic example would be the Linnean classification system, which organizes taxonomic terms into a hierarchy that nests more specific terms within broader ones. In contrast, a vocabulary does not, on its own, achieve any such definition of the relationships between terms.

*Ontologies* and *thesauri* are often used interchangeably to refer to a vocabulary that also includes established relationships between the terms, despite that, semantically, they differ. Compared to vocabularies, thesauri contain a more prominent relationship structure by organizing terms into a hierarchy. Ontologies contain an even greater focus on knowledge representation and provide semantic linkages between vocabulary terms by specifying relationships between terms, which can extend beyond hierarchical to, for example, interspecies interactions (e.g., Wohner et al. 2022). Ontologies were developed to provide greater structure than could any singular vocabulary, thereby making data more accessible, and standardizing across vocabularies (Mi & Thomas 2011, Kartika et al. 2022). Vocabularies can be used for multiple different ontologies, so that ontologies of broadly similar topics (e.g., biology, medicine and ecology) may partially overlap (Kartika et al. 2022). If vocabularies are like a dictionary for data, then ontologies are like the language of data, consisting of not only the words, but also relationships between words and their meanings. Ontologies, while ultimately more helpful for data integration and standards, are also more complex to develop and, therefore, less applied in practice than are vocabularies (Gadelha et al. 2020).

A gradation in complexity can be used, as here, to describe the basics of vocabularies, thesauri and ontologies. However, the standards can also be viewed alongside a continuum which describes a transition from machine actionability to machine readability. For computation, the managed terms of vocabularies provide tagging information that support data aggregation and information retrieval. In contrast, thesauri and, especially, ontologies are designed for machine readable knowledge. They both allow cross-referencing and semantic interfacing with computers (Walls et al. 2014, Schneider et al. 2019, Kartika et al. 2022).

Applications of the *resource description framework* (*RDF*) are, essentially, the computational voices for how data standards can be interfaced with computers. RDF can be used to computationally link two terms (e.g., variables) to reflect relationships, thereby allowing large databases to push away from flat file storage formats (such as GBIF's star schema; Wieczorek et al. 2012, Baskauf et al. 2016, Baskauf & Webb 2016). RDF is, in other words, how basic sentences are computationally formed, explaining, for example, taxonomic hierarchies and species interactions. It explains relationships in a triplet sequence of subject, predicate and object. RDF can computationally express ontological relationships, with the Web Ontology Language, OWL, an example of a data modelling language of data standards that are expressed in RDF.

Despite the many available ontologies to date, the ability to match data to existing sources can remain very low due to the diversity of variables possible to quantify (e.g., Kartika et al. 2022, Wohner et al. 2020, Wohner et al. 2022). A pragmatic approach to implementing data standardization concerns *semantic entity mapping*, which connects and unifies, such as similar terms (variables) between different semantic artifacts (vocabularies, thesauri, ontologies) used to annotate different datasets. Data users may, without realizing it, routinely employ a simple version of mapping when they connect variables when integrating independently produced datasets. They would sort and combine, when possible, the geographical and date variables, or species taxonomy, between different datasets. Mapping can be fairly intuitive to integrate and to synonymize variables between a small number of related datasets, however, it becomes more challenging as datasets increase in size and diversity. It is also subjective to user interpretation. Another potential issue is that it can be employed only so far as there is availability of established data standards for the terms to be mapped, else the decisions become even more subjective (Wieczorek et al. 2012, Walls et al. 2014).

Some argue that ontologies, which reflect interrelationships, should be built to comprehensively cover biodiversity data rather than to use what could be a simpler, but more repetitive and subjective approach to independently map data. The latter "reinvents the wheel" with every mapping occasion, as opposed to the former, which gives instructions on how to build the wheel. However, the latter is much easier to implement directly in data integration, unlike the former, which takes substantial additional work to create the ontologies for the terms of interest, which are nearly as diverse as biodiversity itself. Therefore, mapping is often a more pragmatic solution to data integration than is the development of a specific ontology.

With regard to dataset record-level data standards, there are four important points to keep in mind. First, data standards refer to any individual or collection of vocabularies, thesauri, and/or ontologies (Box 1). They grade in the degree of complexity that they define the relationships between terms, from the simplest, a vocabulary (which lacks relationships), to the more complex, an ontology (which specifies relationships). All can be applied as a standard. Each degree of complexity increases the accuracy of computationally representing the terms (e.g., variables) and their relationships (e.g., taxonomic hierarchy) and, hence, the biodiversity data. For example, a vocabulary can be a data standard, and an ontology is also a data standard. Either may be applied to any given database or data infrastructure, but only one will infer defined relationships between terms.

There is also a hierarchy to data standards, which is important to bring up due to its impact on data interoperability (e.g., computational communication). Data standards vary on how broadly or specifically applicable they are to, and across, disciplines (also referred to as *domains*). Multiple ontologies can sync across disciplines by linking into a *top-level ontology*, which spans across disciplines (Box 1). Top-level ontologies ensure uniform meaning of objects (such as terms) and how they interrelate, irrespective of the discipline they are applicable for. In the early 2000s, the Open Biomedical Ontologies (OBO) consortium formed the OBO-Foundry, which does exactly this for biological and related disciplines (Smith et al. 2007). There is only one level of hierarchy above the OBO-Foundry, and that is the Basic Formal Ontology (BFO), which lacks any domain-specific terminology (Otte et al. 2022). The domain-specific terms must be referenced within the OBO-Foundry and, even more specifically, in its more specialized ontologies. The Biological Collections Ontology (BCO) has started the process to integrate some of the first key terms from Darwin Core in an OBO-Foundry ontology (Walls et al. 2014). Thus, data standards segway from very broad to very specific, and the more discipline-specific standards should, ideally, fit into this hierarchy. The greater relatability discipline-specific ontologies have to top-level ontologies, the better for data interoperability.

The third point is that data standards can be applied, either in part or in full, to any number of databases and data infrastructures. For example, the Darwin Core data standard is a vocabulary that is expressed in RDF (Baskauf et al. 2016, Baskauf & Webb 2016). It is applied as ratified in TDWG for GBIF data (Wieczorek et al. 2012), but it is modified in another data infrastructure, OBIS (the Ocean Biogeographic Information System; De Pooter et al. 2017), to include an extension of the terms covered (see Section 4 for more about *extensions*).

The final point is with respect to ratification. TDWG supports and ratifies many standards, including Darwin Core, but many other vocabularies, thesauri and ontologies also independently exist (Kartika et al. 2022). There can be greater or lesser adoption of them by their scientific communities, i.e., not all are ratified, and even more are being developed on an as-needed basis (Schneider et al. 2019, Gallagher et al. 2020). Data standards are usually created when needed, thus, they rarely follow a single given protocol or method for development, and rarely encompass all terms an individual research project may contain in its dataset. Data standards are community-driven in creation, and TDWG is open for anyone to join. Thus, data standards impose restrictions on data quality and formats (Section 4), but they in themselves are not restricted in terms of who creates them.

## 4. How and whether to select a standard core?

To this point, a straightforward approach to selecting which data can be deposited into a database is to delimit a standard core of common terms (e.g., variables). This approach is employed when centralizing data into a hub that manages it, and requires that dataset record-level data standards have already been developed, to form the standard core. A classic example is how GBIF is structured around a set of Darwin Core standard terms (Wieczorek et al. 2012). Data deposition is streamlined to GBIF by ensuring all datasets contain the core standard variables; however, it also removes or archives extraneous variables which may have originally been a part of the datasets (De Pooter et al. 2017, Guralnick et al. 2018, Gallagher et al. 2020). Information is lost when not all data are included. Standard cores have been created due to the variety of dataset variables; the cores represent the most reliable and common terms (variables) across a variety of datasets, for a given purpose or objective (e.g., museums; Graham et al. 2004), but this cannot cover the whole of biodiversity data. An emerging issue we are facing, as community-based databases and data infrastructures become more prevalent, is the need to include what have been earlier excluded data terms (variables). We provide three options for this, discussing how it may not always be possible to align all data around a core standard that centralizes the data architecture.

One choice is at the level of the data standards governance, whether ratified in TDWG or maintained independently elsewhere. The number of terms in a standard core can be increased through an update of the existing standard core. For example, the ratified TDWG Darwin Core standard began with 24 terms (i.e., variables) in 1998, grew to 169 by their 2009 version (Wieczorek et al. 2012; Darwin Core Maintenance Group 2021) and currently contains 206 terms in their July 2023 version (https://github.com/tdwg/dwc/releases). Still, 206 terms cannot describe all of biodiversity data, and the terms need to be ratified in an update in order to then allow their inclusion in the core. In addition, for any given database or data infrastructure which began based on an older version of a data standard, there will need to be a new solution to how to update to the new standard. For example, they would have to accept blanks in the data deposited before the update for the newly added terms, especially when the data did not originally contain any of the updated terms. Another option could be to retain an archival version of more complete data that can be used to add in at least some of the new terms. Original data sources could also update the data to include variables earlier excluded, if the data managers were motivated and able to. A final option would be to not accept the new version of the standard, and instead maintain data with an older standards version. In the case of GBIF, for such purposes they have made Darwin Core backwards compatible, to account for data updates based on core terms updates. While updating a data standard is very helpful for future databases and data infrastructures, it complicates matters for existing ones, and brings forward a degree of catch-up work for them.

A second option does not need to be directly enacted at the level of data standards governance; it does not modify a ratified data standard. It is applied at the level of an individual database or data infrastructure, and works by including a new optional set of data standards to the existing database, which is referred to as an *extension*, or an *extended core*. When a standard core has already been implemented, which has steered the database structure, and it cannot be drastically modified, this is a practical solution to extend out the data beyond the standard core. Adding an extension of the standard core (which will be a new set of standards) allows data to be integrated that fit with not only the original standard core, but also the newly established extended core. The Humboldt Core (which is ratified by TDWG) is an extension now available for GBIF-deposited data. It allows users to include field inventory information additional to the standard Darwin Core (Guralnick et al. 2018). GBIF contains multiple other extensions (Robertson et al. 2014). Similar logistical issues regarding the implementation of extensions for databases or data infrastructures with earlier deposited data also exist with this option. Usually, the extended core is not required, perhaps somewhat alleviating the issues of adopting one after the fact, at least in practical terms of earlier deposited data. Otherwise, the same issues remain for updating already established databases and data infrastructures as explained for the first option.

The third option is to alter how new databases and data infrastructures are structured for data deposition. By switching to a *decentralized network architecture*, data standards are not required prior to data deposition. They instead may be created after the fact, on an as-needed basis. It bypasses the above issues regarding creating and/or applying only established data standards. A decentralized (but still connected) network architecture is promoted for cases when data standards do not yet exist for the data that are to be integrated. For example, the Open Traits Network is advocating this approach to incentivize community collaboration in integrating diverse traits data that lack any current data standards (Gallagher et al. 2020). It works opposite to the centralized (and connected) network architecture, as is used, for example, with GBIF and OBIS. Considering that ratified data standards take years to be created, removing that requirement would open the potential to integrate a much greater amount of biodiversity data. In theory, the approach promotes data standards creation at a level that supersedes the community-driven approaches of TDWG and similar groups. However, the degree to which such a design really will deviate from the TDWG governance approach to data standards is questionable, i.e., could it really be a current-day reenactment of what originally led to TDWG and the existing data standards governance? How will mapping be implemented to join up similar dataset variables? We caution that a distributed collaboration approach brings with it substantial logistical challenges in terms of community representation and data management (Shaw & Hargittai 2018, O'Leary et al. 2020), but highlight it as an option being promoted by some for breaking out of how they perceive the current data standards regulations to be (Gallagher et al. 2020).

It should be clarified that other bottom-up community approaches have already been tested by both GBIF and TDWG, to support anyone in describing a new term needed for their data. In a recent approach, GBIF is computationally compiling all terms used in datasets that are submitted to and published in GBIF. Many datasets contain more variables than what is ultimately distributed by GBIF in its core and extensions, and these data are available to access in their archives. The inventory of all data terms will be used as a starting point for future discussions to standardize more terms than currently found in controlled vocabularies.

Data standards, however created, are critical for integrating data; otherwise, database contents (i.e., variables) may overlap, not match up properly or be incorrectly defined (i.e., *semantic mismatches*). Data standards also can constrain the extent of data able to be integrated, and thus are not always adopted in fully ratified ways. No approach to working with data integration is a "silver bullet" for biodiversity data, nor can any network structure promise perfect harmonization.

## 5. Case study part I: introducing the Biodiversity Digital Twin and available data standards

*Digital twins* (*DTs*) produce real-time monitoring and decision-making information through the automated processing of data, models and output. As with any modeling approach, DTs require data. They utilize both existing and, more uniquely, real-time sensor data to run models and produce results for people to interpret, whether for industrial, engineering, or, more recently, biodiversity purposes (de Koning et al. 2023, Trantas et al. 2023). The data used by the DT is constantly updated with new information that the DT accesses, then uses to reprocess the modeling output. The revisions happen continuously and, ideally, in real-time. One example of a DT is a weather app. Weather predictions are based on sensor data continuously being fed into weather models, with each being updated continuously to predict current and future conditions. Results are graphed and tabulated, and fed into apps that users view to monitor and check the weather forecasts. The *Biodiversity Digital Twin* (*BioDT*) project has been developed as a DT analog for biodiversity projects - an ambitious endeavor to tackle the world's biodiversity issues in real-time analyses. The remarkability of a BioDT is in the data integration plus modelling – it is a "twinning process" that will allow us to attempt to better understand biodiversity through real-time monitoring as well as predictive modelling for current and future time.

The ambitions do not stop with creating the BioDT, however. Multiple DTs are planned to be integrated into a global system that can represent current to future aspects of the natural, physical and social world, i.e., a computational rendering of earth, and into the future. The BioDT should, thus, be built to be able to connect with other DTs. For example, to interact with the Destination Earth DT, a European Union initiative on climatological, meteorological and atmospheric conditions of the world (Hoffmann et al.

2023). For the DTs to interact, they will require interoperability, and this requires uniform data standards to allow the DTs to commuicate. It is supported through the advancement of open science, such as what the European Open Science Cloud (EOSC) Association is concerned with.

In the BioDT project, ten *prototype Digital Twins* (*pDTs*) have been developed which encompass a variety of biodiversity research topics (Golivets et al. 2024). They are organized around four themes: species and environmental change; genetic biodiversity; threats of policy concern; species and humans. Each pDT requires input data to run their models, but the sources and types of data vary. For example, the pDT data differ in the organismal groups, the taxonomic coverage, the spatiotemporal scales, and the sources of the data, i.e., from the data users' perspectives, they differ in the explanatory and response variables, and where to find those data. The variety in datasets is a challenge for FAIR data and the application of data standards by the BioDT. It also exemplifies the practicalities of integrating biodiversity across an array of topics, data sources, and data types.

The BioDT is pillared by participating *Research Infrastructures* (*RIs*), which are organizations involved in maintaining, integrating and disseminating biodiversity data openly in Europe and/or globally; some are also data infrastructures (i.e., research data infrastructures). The BioDT RIs consist of: GBIF; the Distributed System of Scientific Collections (DiSSCo); the Integrated European Long-Term Ecosystem, critical zone and socio-ecological system Research Infrastructure (eLTER); the e-Science European Infrastructure for Biodiversity and Ecosystem Research (LifeWatch ERIC); and, informally included later in the project, the European life science infrastructure for biological information (ELIXIR). The BioDT RIs differ in their objectives, any forms of data which they integrate and manage, and any applications of data standards (Table 1). The data standards that each utilizes differ, which governs how standardized data may be integrated into the BioDT.

As the RIs are European to global in scope, they are as relevant for other data integration initiatives as for the BioDT project. Whether depositing data into or extracting data from the RIs, how they standardize their data will impact decisions and logistics of data use and further integration (e.g., Wohner et al. 2020). Ideally, all data would filter first through one of the RIs. However, in practice, they do not. The first step to integrating data that will include contributions from the listed RIs, or any further similar organizations, is to understand what data standards they apply, and how the data network is organized (Table 1). In terms of DTs, and given their novelty at this point in time, the RIs require our guidance in how they can better support digital twinning, which includes the need to harmonize their data standards across disciplines (domains) and subdisciplines.

Between the five RIs, GBIF and DiSSCo overlap the most in data standards, because they both deal with museum data (Table 1). They employ the *Access to Biological Collection Data* (*ABCD*) and *Darwin Core* (*DwC*) data standards (Holetschek et al. 2012, Wieczorek et al. 2012, Walls et al. 2014). The ABCD standard defines relationships between terms specific to both specimens (e.g., taxon) and collections (e.g., holding institution). In contrast, DwC terms include those specific to records (e.g., type, basis of record), occurrences (e.g., recorded by), event (e.g., year, month, day), location (e.g., decimal latitude, decimal longitude, altitude), and 13 other categories of terms. DiSSCo independently also utilizes the *Collection Descriptions* (*CD*) for entire collections of natural history museums (e.g., expeditions), and the *Minimum Information about a Digital Specimen* (*MIDS*) for clarifications on digitization. Unique to GBIF is their recent adoption of the *Humboldt Core Extension* (*HumbExt*; https://eco.tdwg.org) to standardize aspects of field inventories (Guralnick et al. 2018). The HumbExt is a vocabulary of field-based terms that, for example, allows the recording of species absences, details on the event (e.g., event duration), sampling effort (e.g., if described and where described, such as in a publication), and similar relevant information not captured by DwC. Note extensions are not required, but possible to be utilized, i.e., not all GBIF data will have data within the HumbExt core.

Data that are deposited within eLTER can contain a diversity of variables that often are difficult to integrate or standardize together, in contrast to the more centralized organization of GBIF and DiSSCo data systems. Instead of a primary core, eLTER data are independent datasets that cover atmospheric, ecological, geological, hydrological & social research topics. The eLTER network architecture is, in other words, more decentralized. As a result, eLTER primarily focuses on standardizing dataset measurements and protocols through their Standards Observations (eLTER SOs; Table 1). For ecological data, eLTER does also support the EnvThes, an ecological thesaurus built on existing vocabularies (Schentz et al. 2013). Examples of EnvThes terms include definitions of species abundance, biomass and growth. However, to our knowledge there is no strict requirement to adhere to EnvThes vocabulary, nor does EnvThes match to any of the standards applied by GBIF and DiSSCo. eLTER and EnvThes focus on site-level field data (Wohner et al. 2022).

The final two RIs, LifeWatch ERIC and ELIXIR, are not directly involved with data aggregation, and so have not adopted any specific data standards. LifeWatch ERIC primarily facilitates biodiversity data logistics. Among its resources is the EcoPortal ([https://ecoportal.lifewatch.eu](https://ecoportal.lifewatch.eu)), a repository of about 30 ecological data standards and related semantic domains, including EnvThes (supported by eLTER). ELIXIR's focus is more on life science data, i.e., biomolecular and chemically-derived data. They also provide guidance and weblinks for databases and standards help. As with LifeWatch ERIC, the ELIXIR reference lists can be a stepping-stone towards finding relevant data sources and/or standards.

If all data will originate from within an RI, any research project requiring the data can bypass further data standards applications. However, if data originate from multiple RIs and/or only partly from an RI, the data must be standardized to be harmonized. Six of the ten BioDT pDT projects utilize biodiversity data from either GBIF or other independent sources (i.e., otherwise published or private), and two utilize eLTER data (Figure 1). The differences between the data standards of GBIF and eLTER will pose challenging for any direct data integration procedures between them.

**6. Case study part II: the ways to standardize biodiversity data during integration for complex research and DTs.**

Philosophies on data standards range from the most idealistic, top-level approaches that utilize and contribute to standards that are suitable across all possible disciplines (domains), to the most practical, applied methods that directly address one specific data integration case without relevance to even their own discipline within the field of data standards. There is a struggle between motivating the field of data standards forward against a lack of knowledge about the existence of such a field or relevance to individual researchers, combined with the weight each individual research project bears to complete their task(s) in a limited and timely manner.

In the case of the BioDT pDTs, and many data users' objectives, data are created and/or selected for the research, and not based on existing standards or data infrastructures. The goals are to integrate the data, but only to utilize them in modelling. However, the more data standards can be advanced alongside the actual research, the more it will aid future data standardization requirements. In this final section, we explain the possible ways to standardize biodiversity data during integration, transitioning from ideal to practical. The efforts and outcomes are a balance between practical constraints in enacting standardizations versus the advancement of biodiversity data standards as a field. The labels we provide for each approach are not meant to serve as lasting names, but rather to help differentiate between the other possible options we propose.

The first approach to data standards is the most broadly applicable. It is conducted with relevance across all domains, by basing a new biodiversity ontology on a top-level ontology. We thus call this the "*top-ontology*" approach to data integration, as it begins most broadly, for example at the level of the OBO-Foundry (Section 3), to ensure all biodiversity data objects (such as terms) will be interoperable across the lower-level ontologies and different disciplines. In such a case, data and information scientists could work with RIs to ensure and guide them in their data standards efforts, thereby producing an upper-level ontology with interoperability capable of, for example, linking different DTs.

In the next approach, which we dub the "*single-ontology*" approach, would be to build off of existing ontologies to unify them, and to define the relationships between all of the data terms (variables) of the different biodiversity datasets. Like the "top-ontology approach," it would advance the biodiversity data standards field, and in such a way result in less repetition for future initiatives. Communication with the general data standards community would be critical, for awareness as well as for developing the ontology. The ontology would be built from established data standards, not starting from scratch, to add in terms not yet established. One by one, a list of all possible terms would need to be compiled, mapped to existing vocabulary and ontologies, and then filled in with further information to create a full biodiversity ontology. For example, in the case of the BioDT, it would be built from ABCD, DwC, eLTER SO, EnvThes, and HumbExt data standards, applying them to the different data sources of the pDTs (Table 1). Semantic mismatches would need to be identified, alongside establishing terms, the definitions, synonym terms, and the interrelationships between terms. Another similar, sub-approach would stop at building a comprehensive vocabulary, instead of an ontology, and thus be somewhat more achievable. Relationships between terms would not need to be established, but the bulk of the work in defining terms and, potentially, synonyms would still be required.

The efforts of both an "top-ontology" and "single-ontology" approaches to integration would be substantial, and most practical if the general data standards community (i.e., TDWG) and the RIs could drive it forward. To succeed, it would require extensive time for discussion and implementation; more than is afforded in most research project durations. The process itself would also require a community component and could bring up a myriad of conversations and opinions on the semantics of terms. How practical such an approach is within the realism of funding timing is questioned. However, it can alternatively be argued that even beginning such a task, to allow then the TDWG community and/or RIs to take it over, would be substantial progress in the field of data standards. For linking DTs, to do so without having any standards based on top-level ontologies would prove an impossible endeavor, with very little progress likely capable of being made towards the vision of a Destination Earth unifying the DTs. Even given how diverse the BioDT pDT projects are (Golivets et al. 2024), our goal is more to guide the direction of data standards (especially within and among RIs) towards interoperability. Within the timespan of the BioDT, it would be nearly impossible to build any unifying ontology or vocabulary to cover the data of all ten pDTs.

In similar situations, one bottom-level variation of the "single-ontology" approach could be to focus on one topic (e.g., crop ontologies or another topic of the pDTs for the BioDT). Usually, it would connect to a researcher's own interests, and while it would help move forward the data standards field, it would not suffice to meet a project's needs when data standards must cover the whole of a variety of topics, nor would it be cross-domain. It would likely still be a daunting task for someone to begin who has not already gained knowledge in the data standards field (hence our guide here). By developing research projects which collaborate with a data scientist's experience in data standards, the "single-ontology" approach should become more feasible to implement with time and sustained investment. This is an approach, for example advocated but not implemented in a review of biodiversity digital twin challenges (Trantas et al. 2023).

The third approach is more pragmatic to implement, but it also will not forward the state of biodiversity data standards to such degrees as the prior two. In what we call the "*map to existing approach*", it begins similar to the prior two methods. The first step would be to find any applicable and already established data standards, for example as we have done here for the BioDT (Table 1). The existing data standards could be compiled, as possible, into a list of terms that are then used to map the research data terms (variables) to. If research data terms do not match to an existing standard, they should still be matched, as relevant, between the dataset variables of the project. In such an approach, there would be no effort to create new data standards, but existing applicable standards would be utilized to map the research data. In the same way that lists can help notify the scientific community of available datasets (i.e., Section 2), an initiative which compiled existing biodiversity data standards would be helpful for the data standards community. This approach has been implemented by Wohner et al. (2020) to create a set of core site-based terms, analogous to the Darwin Core of GBIF, with additional recommended terms for lesser used field terms. Other RIs such as LifeWatch ERIC and ELIXIR have begun similarly, but stopped short at supplying links to the different vocabularies. We suggest that the "map to existing" approach is better suited to research projects that do not cover a large extent of biodiversity topics, as otherwise, and as with the first two approaches, the implementation becomes challenging within the full data scope, as for example, with the BioDT.

The final alternative, which we call the "*map independently*" approach, is the easiest to apply, but it also fails to forward the field of data standards. Independently, without regard to any existing data standards, data can be mapped between the different sources. It will achieve the goals of a research project to be able to model the integrated data. It is the method that, essentially, researchers utilize by default when independently integrating data. The lack of data standards advancement alongside high subjectivity are cons against the timeliness and ease to enact this approach. In the case of the BioDT, the interdisciplinary nature of the highly varied biodiversity data of the pDTs has prompted us to adopt the "map independently" approach. It should be noted that over half of the pDTs do utilize an RI for data, hence, much of the data terms are already standardized. In alignment with EOSC's Metadata Schema and Crosswalk Registry, the BioDT project is looking into mapping the pDT data through a newly developed Mapping.bio tool (Wolodkin et al. 2023). We recognize the importance of data standards and furthering the field, as we have taken care to explain throughout this guide. Ultimately, however, we are as tethered to fulfilling the project goals as is the next research project. It demonstrates a conundrum to data standards: how do we advance a field that itself is all too often an "after thought" to a methodological component of a research project? We cannot fully answer our question, but can begin to by raising awareness of the issue through the production of this guide.

## 7. Conclusion.

Data standardization is an instrument for the integration of massively diverse data, as needed for complex research projects such as the BioDT. Mismatches between data standards of existing biodiversity research data infrastructures render it challenging to adhere fully to them across clustered communities of data producers and data users. Research infrastructures will continue to augment their data contents, which may make them increasingly more relevant for data users. The challenges facing data standardization by an independent research project may be too many for most to overcome, but projects can at least be aware of and adhere to the existing standards that have been implemented for their data utilized in a project. Much of the work with data standards will be achieved within research infrastructures, alongside inter-research collaboration and wider open-science initiatives like EOSC and Destination Earth, and, finally, with interoperability from cross-digital twin project collaborations.

**Acknowledgements.**

Dmitry Schigel is thanked for contributions to our knowledge of the data sources of the BioDT projects, and for helpful manuscript edits and comments. This project has received funding from the European Union's Horizon Europe research and innovation programme under grant agreement No 101057437 (BioDT project, https://doi.org/10.3030/101057437). Views and opinions expressed are those of the author(s) only and do not necessarily reflect those of the European Union or the European Commission. Neither the European Union nor the European Commission can be held responsible for them.

**Declaration of competing interest.**

There are no conflicts of interest that need to be declared.

**Author contributions.**

C. Andrew: Conceptualization, Data – BioDT pDT projects, Visualization, Writing – original draft, Writing – review & editing. S. Islam: Data – BioDT pDT projects, Writing – review and editing. C. Weiland & D. Endresen: Writing – review and editing.

**Data availability.**

Further information on the BioDT pDT projects are published in the special issue: Golivets M, Sharif I, Wohner C, Grimm V, & Schigel D (eds). 2024. Building Biodiversity Digital Twins. RIO. [https://doi.org/10.3897/rio.coll.240](https://doi.org/10.3897/rio.coll.240). No other data are referenced.

**Figures and boxes (with captions)**.

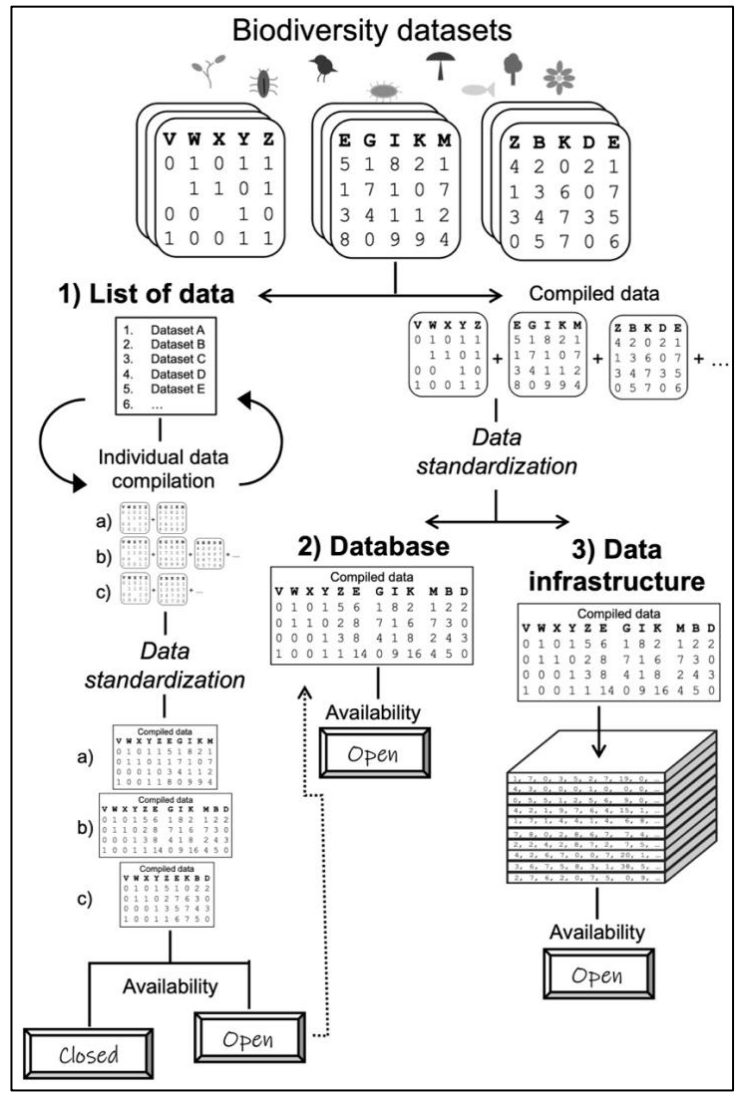

Figure 1. Data production flow chart. To make biodiversity datasets available, they must first be made interoperable through shared data standards. Data standardization is critical to compilation of data, whether from lists of available datasets, singular databases, or data within data infrastructures. Despite the criticality of data standardization, the process is rarely discussed, except at institutional levels of data infrastructures.

> **Box 1. There are many standards.**
>
> A) **Data standards:** consistent specifications for expressing data for storage & exchange
>
> B) **Types of data standards vary**
>   - **Parameter-level**
>     - Specific to units of measure & values
>     - International Organization for Standardization (ISO):
>       - Internationally-defined standards for country codes, currency, date and time, food safety, IT, language, & more
>     - Examples: grams, Celsius, YYYY-MM-DD
>   - **Dataset-level**
>     - Related to aspects of dataset structure, discipline (domain), terms (variables), & relationships
>     - For between-dataset integration & computational accessibility
>       - **Dataset record-level**
>         - Focused on terms (variables) & their relationships
>         - Includes *vocabularies*, *thesauri* & *ontologies*
>         - Biodiversity Information Standards (TDWG): https://www.tdwg.org
>   - **Computational**
>     - Metadata standards
>       - Describe data to computers via data structures in schemas
>       - Formulated via standardized languages
>
> C) *More about dataset record-level standards*
>   - **Semantic artifacts / entities**
>     - Broader term to include *vocabularies*, *thesauri*, *ontologies*, metadata schemes, terminologies, & taxonomies
>   - **Vocabularies**
>     - Lists of terms with definitions
>     - Organize things (e.g., terms) into an inventory
>     - Like a dictionary
>   - **Thesauri**
>     - Adds to a vocabulary
>     - Organizes things (e.g., terms) into a hierarchical relationship structure
>   - **Ontologies**
>     - Focus on knowledge representation
>     - Semantically links vocabulary terms
>     - Specified relationships can be more than hierarchical (e.g., interspecies interactions)
>     - Like a language
>       - **Ontological hierarchies**
>         - Top-level: span across disciplines
>         - OBO-Foundry for biological & related disciplines: https://obofoundry.org
>         - Basic Formal Ontology (BFO) lacks all domain-specific terminology: https://basic-formal-ontology.org
>   - **Resource description Framework**
>     - A computational language model to express semantic artifacts
>     - Expressed in Web Ontology Language (OWL)

Box 1. Explanations of what data standards are, and comparisons of important terminologies and relationships between the standards.

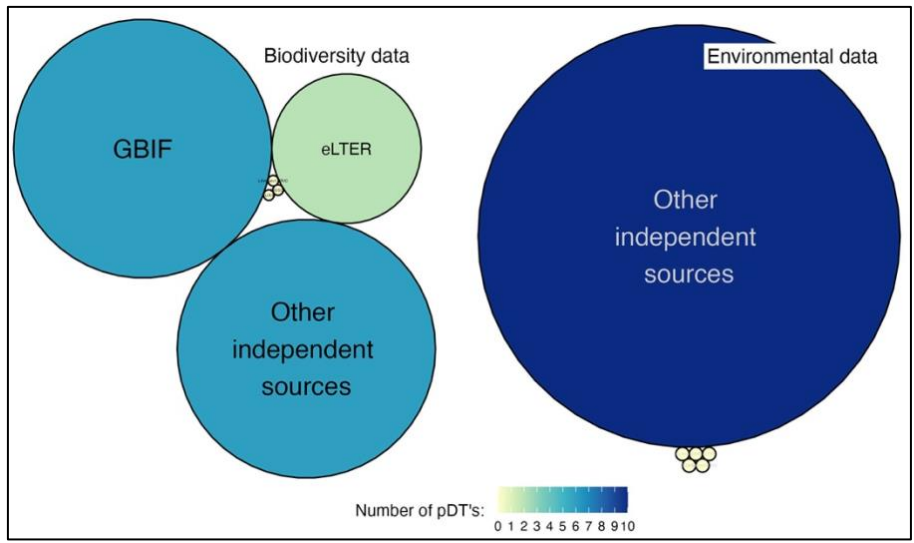

Figure 2. Number of BioDT pDTs utilizing a Research Infrastructure or other source(s) for their input data. The source(s) that the pDTs utilize depend, in large extent, on the types of data. Biodiversity data are used to model biological aspects of biodiversity (i.e., what is often predicted; left side), while environmental data are used to predict responses (i.e., what are often explanatory data; right side). Circles are sized and shaded by the number of pDTs utilizing an RI or other source. There are a total of 10 pDTs.

**Tables (with captions)**.

Table 1. Currently published data standards used by the different BioDT Research Infrastructures (GBIF, LifeWatch ERIC, DiSSCo, eLTER & ELIXIR). The research infrastructures contribute differently towards aspects of biodiversity data. The Research Infrastructures and data standards link to webpages of each. The data standards most applicable to biodiversity data are distinguished by a larger, bolder X.

| Research infrastructures | | | | Data standard(s) applied to data | | | | | | |
|---|---|---|---|---|---|---|---|---|---|---|
| **Name** | **Acronym** | **Year initiated** | **Type of contribution to biodiversity data** | **ABCD** | **CD** | **DwC** | **eLTER SO** | **EnvThes** | **HumbExt** | **MIDS** |
| Global Biodiversity Information Facility | GBIF | 2001 | Occurrence, checklist & sampling-event data | X | | X | | | X | |
| European life science infrastructure for biological information | ELIXIR | 2013 | Life science resources & interoperability services (storage, access, analyses) | *Not applicable (provides links to repositories of established standards)* | | | | | | |
| e-Science European Infrastructure for Biodiversity and Ecosystem Research | LifeWatch ERIC | 2017 | e-Science research facilities & services (Virtual Research Environments) | *Not applicable (provides links to repositories of established standards; EcoPortal)* | | | | | | |
| Distributed System of Scientific Collections | DiSSCo | 2018 | Natural science (museum) collections data | X | X | X | | | | X |
| Integrated European Long-Term Ecosystem, critical zone and socio-ecological system Research Infrastructure | eLTER | 2018 | Integrated (biotic, abiotic & social) datasets | | | | X | X | | |

*Year initiated: For DiSSCo & eLTER, this refers to when they joined the ESFRI Roadmap. For LifeWatch ERIC, this refers to their EU establishment. For GBIF, this refers to its official establishment. For ELIXIR, this refers to its permanent phase (the preparatory phase began in 2007)